\documentclass[draft,12pt]{article}
\usepackage{amsmath,amsfonts,palatino,amsthm,epsf}
\usepackage{cite}
\newcommand{\beq}{\begin{equation}}  
\newcommand{\eeq}{\end{equation}}

\begin{document}

\title{Electroweak and Majorana Sector Higgs Bosons and Pseudo-Nambu-Goldstone Bosons}
\author{Wei Lu
\thanks{New York, USA, email address: weiluphys@yahoo.com}
\\
\\
}
\maketitle

\begin{abstract}
We propose a Clifford algebra based model, which treats both gravity and Yang-Mills interactions as gauge fields.  There are two sectors of boson fields as electroweak and Majorana bosons. The electroweak boson sector induces fermion masses via spontaneous symmetry breaking. It is composed of scalar Higgs, pseudoscalar Higgs, and antisymmetric tensor components. The Majorana boson sector contributes to flavor mixing and Majorana masses of right-handed neutrinos. It is comprised of neutrino Higgs and pseudo-Nambu-Goldstone bosons. The LHC 750 GeV diphoton resonance might possibly be identified as a Majorana sector pseudo-Nambu-Goldstone boson, which results from spontaneous symmetry breaking of a flavor-related global $U(1)$ symmetry involving four-fermion condensation of right-handed leptons and quarks. The diphoton decay is loop induced, since tree-level decay is suppressed by large Majorana mass of the right-handed neutrino. There is also a potential dark matter candidate, which is the four-lepton condensation of muon, muon-neutrino, tau, and tau-neutrino.
\end{abstract}

{\bf Keywords}. Higgs Bosons, Pseudo-Nambu-Goldstone Bosons.

\newpage

\section{Introduction}

Clifford algebra, also known as geometric algebra or space-time algebra, has found a wide variety of applications in physics\cite{HEST1,HEST2,CHIS,BORS,TRAY,BAYL,BESP,PAVS, LOUN,BUDI,DORA,RODR,NEST,BOUD,WL1,DAVI,CIRI}. Attempts have been made to identify species of fermions as ideals (idempotent projections of the original spinor) and derive Standard Model gauge symmetries from various dimensions of Clifford algebras. 

We propose a model which is based on Clifford algebra $C\!\ell_{0,6} \oplus C\!\ell_{T1}$. It includes local gauge symmetries $SO(1,3)_{LOR} \otimes SU(2)_{WL} \otimes U(1)_{WR} \otimes U(1)_{B-L} \otimes SU(3)_C$. There are two sectors of bosonic fields as electroweak and Majorana bosons. The electroweak sector induces fermion masses via spontaneous symmetry breaking. The Majorana sector contributes to flavor mixing and Majorana masses of right-handed neutrinos. 

The experiments at LHC recently indicated a diphoton resonance at about 750 Gev\cite{H750A, H750C}. The resonance might possibly be identified as a Majorana sector pseudo-Nambu-Goldstone boson, which results from spontaneous symmetry breaking of a flavor-related global $U(1)$ symmetry involving four-fermion condensation of right-handed leptons and quarks. 

This paper is structured as follows: Section 2 introduces gauge symmetries and the gauge-invariant action. In section 3, the Majorana boson sector is discussed. In section 4, we study electroweak boson sector. In section 5, we briefly touch upon the topic of grand unification symmetries. In the last section we draw our conclusions.

\section{Gauge Symmetries and Gauge-Invariant Action}
\subsection{Leptons, Quarks, and Projection Operators}

We begin with a review of Clifford algebra $C\!\ell_{0,6}$\cite{WL1}. It is defined by anticommutators of orthonormal vector basis $\{ \gamma_{j}, \Gamma_{j}; {j}= 1,2,3\}$ 
\begin{align}
&[\gamma_{j}, \gamma_{k}] = \frac{1}{2}(\gamma_{j}\gamma_{k}+\gamma_{k}\gamma_{j})=-\delta_{jk},\\
&[\Gamma_{j}, \Gamma_{k}] =-\delta_{jk},\\
&[\gamma_{j}, \Gamma_{k}] =0,
\end{align}
where ${j}, {k}=1, 2, 3.$ All basis vectors are space-like. There are ${{{{{{{{{{{{{{{{{{{{{\binom{6 }{k}}}}}}}}}}}}}}}}}}}}}} $ independent $k $-vectors. The complete basis for $C\!\ell_{0,6}$ is given by the set of all $k $-vectors. Any multivector can be expressed as a linear combination of $2^{6}=64$ basis elements. 

Two trivectors
\begin{align}
&\gamma_{0} = \Gamma_{1}\Gamma_{2}\Gamma_{3}, \\
&\Gamma_{0} = \gamma_{1}\gamma_{2}\gamma_{3}
\end{align}
square to 1, so they are time-like. The orthonormal vector-trivector basis $\{\gamma_a, a= 0,1,2,3\}$ defines space-time Clifford algebra $C\!\ell_{1,3}$, with 
\begin{equation}
\left\langle \gamma_{a}\gamma_{b}\right\rangle = \eta_{ab} = diag(1, -1, -1, -1),
\end{equation}
where $\left\langle \cdots\right\rangle$ means scalar part of enclosed expression. The reciprocal vectors $\{\gamma^{a}\}$ are defined by 
\begin{equation}
\gamma^{a}\eta_{ab} = \gamma_{b}, 
\end{equation}
thus
\begin{equation}
\left\langle \gamma^{a}\gamma_{b}\right\rangle = \delta^{a}_{b}.
\end{equation}
Here we adopt the summation convention for repeated indices. Notice that $\gamma_{0}$ is a trivector, rather than a vector. 

The unit pseudoscalar
\begin{equation}
i =  \Gamma_{1}\Gamma_{2}\Gamma_{3}\gamma_1\gamma_2\gamma_3=\gamma_0\gamma_{1}\gamma_{2}\gamma_{3}=\gamma_0\Gamma_{0}
\end{equation}
squares to $-1$, anticommutes with odd-grade elements, and commutes with even-grade
elements. 

Reversion of a multivector $M\in C\!\ell_{0,6}$, denoted $\tilde{M}$, reverses the
order in any product of vectors. For any multivectors $M$ and $N$, there are algebraic properties 
\begin{align}
&(MN)^{\tilde{}} = \tilde{N}\tilde{M}, \\
&\left\langle MN\right\rangle =\left\langle NM\right\rangle.
\end{align}
The magnitude of a multivector $M$ is defined as
\begin{equation}
|M| = \sqrt{\left\langle M^{\dagger}M \right\rangle},
\end{equation}
where 
\begin{equation}
M^{\dagger} = -i\tilde{M}i,
\end{equation}
is the Hermitian conjugate.

Algebraic spinor $\psi\in C\!\ell_{0,6}$ is a multivector , which is expressed as a linear combination (with Grassmann-odd coefficients) of all $2^{6}=64$ basis elements. 

Spinors with left/right chirality correspond to odd/even multivectors
\begin{align}
&\psi = \psi_L + \psi_R, \\
&\psi_L = \frac{1}{2}(\psi + i\psi i), \\
&\psi_R = \frac{1}{2}(\psi - i\psi i). 
\end{align}

A projection operator squares to itself. Idempotents are a set of projection operators
\begin{align}
&P_l = \frac{1}{4}(1+iJ_1+iJ_2+iJ_3) = \frac{1}{4}(1+3iJ), \label{IDEM2}\\
&P_{q1} = \frac{1}{4}(1+iJ_1-iJ_2-iJ_3), \label{IDEM3}\\
&P_{q2} = \frac{1}{4}(1-iJ_1+iJ_2-iJ_3), \label{IDEM4}\\
&P_{q3} = \frac{1}{4}(1-iJ_1-iJ_2+iJ_3), \label{IDEM5}\\
&P_q = P_{q1}+P_{q2}+P_{q3} = \frac{3}{4}(1-iJ), \label{IDEM1}\\
&P_{\pm} = \frac{1}{2}(1\pm {\Gamma_0}{\Gamma_3}), \label{IDEM6}
\end{align}
where 
\begin{align}
&J_1 = \gamma_1\Gamma_1, J_2 = \gamma_2\Gamma_2, J_3 = \gamma_3\Gamma_3, \\
&J = \frac{1}{3}(J_1 + J_2 + J_3),\\
&P_l + P_{q1} + P_{q2} + P_{q3} = P_l + P_q = 1, \\
&P_{a}P_{b} = \delta_{ab},  \quad (a, b = l, q1, q2, q3), \\
&P_+ + P_- = 1. 
\end{align}
Here $P_l$ is lepton projection operator, $P_q$ is quark projection operator, and $P_{qj}$ are color projection operators. The bivectors $J_j$ appearing in the color projectors $P_{qj}$ suggest an interesting duality between 3 space dimensions and 3 colors of quarks.

Now we are ready to identify idempotent projections of spinor
\begin{equation}
\psi = (P_+ + P_-)(\psi_L + \psi_R)(P_l + P_{q1} + P_{q2} + P_{q3})
\end{equation}
with left-handed leptons, red, green, and blue quarks
\begin{equation}
\left\{
\begin{array}{rl}
&\nu_L = P_+\psi_LP_l, \\
&e_L = P_-\psi_LP_l, \\
&u_L = P_+\psi_LP_{q1} + P_+\psi_LP_{q2} + P_+\psi_LP_{q3} = P_+\psi_LP_q,  \\
&d_L = P_-\psi_LP_{q1} + P_-\psi_LP_{q2} + P_-\psi_LP_{q3} = P_-\psi_LP_q,
\end{array}\right.
\end{equation}
and right-handed leptons, red, green, and blue quarks
\begin{equation}
\left\{
\begin{array}{rl}
&\nu_R = P_-\psi_RP_l, \\
&e_R = P_+\psi_RP_l, \\
&u_R = P_-\psi_RP_{q1} + P_-\psi_RP_{q2} + P_-\psi_RP_{q3} = P_-\psi_RP_q,  \\
&d_R = P_+\psi_RP_{q1} + P_+\psi_RP_{q2} + P_+\psi_RP_{q3} = P_+\psi_RP_q.\end{array}
\right.
\end{equation}

\subsection{Gauge Symmetries}
Spinors transform as 
\begin{align}
&\psi_L\rightarrow e^{\Theta_{LOR} + \Theta_{WL}}\psi_L e^{\Theta_J  - \Theta_{STR}}, \label{TRANSL}  \\
&\psi_R\rightarrow e^{\Theta_{LOR} + \Theta_{WR}}\psi_R e^{\Theta_J - \Theta_{STR}}. \label{TRANSR}
\end{align}
It is worth noting that all gauge transformations are with Grassmann-even rotation angles, so that the transformed spinors remain to be Grassmann-odd.

There are Lorentz $SO(1,3)_{LOR}$  gauge transformations 
\begin{equation}
\{\gamma_a\gamma_b\} \in \Theta_{LOR}, \label{LOR} \\
(a, b = 0, 1, 2, 3, a \neq b), 
\end{equation}
weak isospin $SU(2)_{WL}$ gauge transformations acting on left-handed fermions 
\begin{equation}
\{\frac{1}{2}\Gamma_2\Gamma_3, \frac{1}{2}\Gamma_1\Gamma_3, \frac{1}{2}\Gamma_1\Gamma_2\} \in\Theta_{WL}, \label{WL}
\end{equation}
weak $U(1)_{WR}$ gauge transformation acting on right-handed fermions 
\begin{equation}
\{\frac{1}{2}\Gamma_1\Gamma_2\} \in\Theta_{WR}, \label{WR}
\end{equation}
$U(1)_{B-L}$ gauge transformation 
\begin{equation}
\{\frac{1}{2}J\} \in \Theta_{B-L}, \label{J}
\end{equation}
and color $SU(3)_C$ gauge transformations
\begin{equation}
\left\{
\begin{array}{rl}
&T_1, T_2, T_3, \\
&T_4, T_5, \\
&T_6, T_7, \\
&T_8
\end{array}
\right\}
= 
\left\{
\begin{array}{rl}
&\frac{1}{4}(\gamma_1\Gamma_2 + \gamma_2\Gamma_1), 
\frac{1}{4}(\Gamma_1\Gamma_2 + \gamma_1\gamma_2), 
\frac{1}{4}(\Gamma_1\gamma_1 - \Gamma_2\gamma_2), \\
&\frac{1}{4}(\gamma_1\Gamma_3 + \gamma_3\Gamma_1),
\frac{1}{4}(\Gamma_1\Gamma_3 + \gamma_1\gamma_3),  \\
&\frac{1}{4}(\gamma_2\Gamma_3 + \gamma_3\Gamma_2),
\frac{1}{4}(\Gamma_2\Gamma_3 + \gamma_2\gamma_3),  \\
&\frac{1}{4\sqrt{3}}(\Gamma_1\gamma_1 + \Gamma_2\gamma_2 - 2\Gamma_3\gamma_3) 
\end{array}
\right\} \quad \in \Theta_{STR}.  \label{SU3}
\end{equation}

Notice that the gauge groups contain both gravitational ($SO(1,3)_{LOR}$)\footnote{ See ref. \cite{LORE, POIN, DESI, GGRAV} for various gauge gravity theories.} and Yang-Mills gauge transformations. 

Because the product of lepton projector $P_l$ with any generator in color algebra (\ref{SU3}) is zero $P_lT_k = 0$, leptons are invariant under color gauge transformations. 

After symmetry breaking of $SU(2)_{WL}$, $U(1)_{WR}$, and $U(1)_{B-L}$ via Majorana and electroweak Higgs bosons, which will be detailed in later sections, the remaining electromagnetic $U(1)$ symmetry is a synchronized double-sided transformation
\begin{align}
&\psi\rightarrow e^{\frac{1}{2}\epsilon_E\Gamma_1\Gamma_2}\psi e^{\frac{1}{2}\epsilon_E J}, \label{EM}
\end{align}
where a shared rotation angle $\epsilon_E$ synchronizes the double-sided gauge transformation. 

Thanks to the properties
\begin{align}
&JP_l = (B-L)iP_l = -iP_l, \label{J0}  \\
&JP_{qj} = (B-L)iP_{qj} = \frac{1}{3}iP_{qj}, \label{Jq}  \\
&\Gamma_1\Gamma_2P_{\pm} = {\mp} iP_{\pm}, \label{G3}
\end{align}
electric charges $q_k$ as in
\begin{align}
&e^{\frac{1}{2}\Gamma_1\Gamma_2}\psi_k e^{\frac{1}{2} J} = \psi_k e^{q_ki} \label{EM2}
\end{align}
are calculated as $q_k = 0, -1, \frac{2}{3}$, and $-\frac{1}{3}$ 
for neutrino, electron, up quarks, and down quarks, respectively. Here $B$ and $L$ are baryon and lepton numbers, respectively. 

\subsection{Gauge Field 1-Forms, Gauge-Covariant Derivatives, and Curvature 2-Forms}

Gauge fields are Clifford-valued 1-forms (Clifforms\cite{CF1, CF2} with Grassmann-even coefficients) on 4-dimensional space-time  manifold ($x_{\mu}$, $\mu = 0, 1, 2, 3$)
\begin{align}
&e = e_{\mu}dx^{\mu} = e^{a}_{\mu}\gamma_{a}dx^{\mu}, \\
&\omega = \omega_{\mu}dx^{\mu} = \frac{1}{4}\omega^{ab}_{\mu}\gamma_{a}\gamma_{b}dx^{\mu} \quad\in\quad \Theta_{LOR}, \\
&W_L = W_{L\mu}dx^{\mu}= \frac{1}{2}(W^1_{L\mu}\Gamma_2\Gamma_3 + W^2_{L\mu}\Gamma_1\Gamma_3 + W^3_{L\mu}\Gamma_1\Gamma_2) dx^{\mu} \quad\in\quad \Theta_{WL},\\
&W_R = W_{R\mu}dx^{\mu}= \frac{1}{2}W^3_{R\mu}\Gamma_1\Gamma_2dx^{\mu} \quad\in\quad \Theta_{WR},\\
&C = C_{\mu}dx^{\mu}= \frac{1}{2}C^J_{\mu}Jdx^{\mu} \quad\in\quad \Theta_{B-L}, \\
&G = G_{\mu}dx^{\mu}= G^k_{\mu}T_kdx^{\mu} \quad\in\quad \Theta_{STR},
\end{align}
where $e$ is vierbein, $\omega$ is gravity spin connection, $G$ is strong interaction, and the rest are electroweak related interactions. Notice that we adopt the same notation for vierbein 
$e$ , mathematical number $e$, and electron $e$. One should be able to differentiate them based on contexts.

The vierbein field $e$ acts like space-time frame field, which is essential in building all actions as diffeomorphism-invariant integration of 4-forms on 4-dimensional space-time manifold. The space-time manifold is initially without metric. It's the vierbein field which gives notion to metric 
\begin{equation}
g_{\mu\nu} = \left\langle e_{\mu}e_{\nu} \right\rangle = e^{a}_{\mu}e^{b}_{\nu}\eta_{ab}.
\end{equation}

Local gauge transformations are coordinate-dependent gauge transformations. Gauge fields obey local gauge transformation laws
\begin{align}
&e(x) \quad\rightarrow\quad  e^{\Theta_{LOR}(x)}e(x)e^{-\Theta_{LOR}(x)}, \\
&\omega(x) \quad\rightarrow\quad  e^{\Theta_{LOR}(x)}\omega(x)e^{-\Theta_{LOR}(x)} - (de^{\Theta_{LOR}(x)})e^{-\Theta_{LOR}(x)}, \\
&W_L(x) \quad\rightarrow\quad  e^{\Theta_{WL}(x)}W_L(x)e^{-\Theta_{WL}(x)} - (de^{\Theta_{WL}(x)})e^{-\Theta_{WL}(x)}, \\
&W_R(x) \quad\rightarrow\quad  W_R(x) - (de^{\Theta_{WR}(x)})e^{-\Theta_{WR}(x)}, \\
&C(x) \quad\rightarrow\quad  C(x) - e^{-\Theta_{B-L}(x)}(de^{\Theta_{B-L}(x)}), \\
&G(x) \quad\rightarrow\quad  e^{\Theta_{STR}(x)}G(x)e^{-\Theta_{STR}(x)} + e^{\Theta_{STR}(x)}(de^{-\Theta_{STR}(x)}),
\end{align}
where $d = dx^{\mu}\partial_{\mu}$.

It's worth emphasizing that gravity related fields $e(x)$ and $\omega(x)$ are treated as gauge fields with local gauge transformation properties, as the rest Yang-Mills gauge fields. 

Gauge-covariant derivatives of spinor fields $\psi_{L/R}(x)$ are defined by
\begin{align}
&D\psi_L = (d + \omega + W_L)\psi_L + \psi_L (C - G), \\
&D\psi_R = (d + \omega + W_R)\psi_R + \psi_R (C - G).
\end{align}
The gravitational spin connection $\omega$ is essential in maintaining {\it local } Lorentz covariance of $D\psi_{L/R}$.

We introduce gauge curvature $2$-forms by applying the covariant derivative to the $0$-form spinor $\psi$ and then to the $1$-form spinor $D\psi$
\begin{align}
D(D\psi_{L/R}) &= (d + \omega + W_{L/R})D\psi_{L/R} - D\psi_{L/R} (C - G) \\
							&= (R + F_{WL/WR}) \psi_{L/R} (F_J - F_{STR}),
\end{align}
where gravity, left/right weak, $J$, and strong force curvature 2-forms are
\begin{align}
&R = d\omega + \omega^2 = \frac{1}{2} R_{\mu\nu} dx^{\mu}dx^{\nu}, \\
&F_{WL} = dW_L + W_L^2 = \frac{1}{2} F_{WL\mu\nu} dx^{\mu}dx^{\nu}, \\
&F_{WR} = dW_R = \frac{1}{2} F_{WR\mu\nu} dx^{\mu}dx^{\nu}, \\
&F_{J} = dC = \frac{1}{2} F_{J\mu\nu} dx^{\mu}dx^{\nu}, \\
&F_{STR} = dG + G^2 = \frac{1}{2} F_{STR\mu\nu} dx^{\mu}dx^{\nu}. 
\end{align}
$F^{\mu\nu k}$ is defined by
\begin{align}
&F^{\mu\nu k}\eta_{\mu\alpha}\eta_{\nu\beta} = F^k_{\alpha\beta},
\end{align}
where $k$ enumerates the Clifford components of each gauge field.

Notice that the connection fields are defined to absorb gauge coupling constants. As a result, gauge coupling constants neither appear in the definition of gauge-covariant derivatives of fermions $D\psi_{L/R}$, nor appear in the gauge curvature $2$-forms such as $F_{WL} = dW_L + W_L^2$. They will show up in the gauge field actions instead.

\subsection{Gauge-Invariant Action}

The local gauge- and diffeomorphism-invariant action is 
\begin{align}
S_{World} =  &S_{Spinor-Kinetic}  \\
+ &S_{Gravity} + S_{Yang-Mills} \\
+ &S_{Majorana-Yukawa} + S_{Majorana-Bosons}  \\
+ &S_{Electroweak-Yukawa} + S_{Electroweak-Bosons}.
\end{align}

The spinor kinetic action is now written down as
\begin{equation}
S_{Spinor-Kinetic} \sim \int{\left\langle \bar{\psi}_Lie^3 D\psi_L + \bar{\psi}_Rie^3 D\psi_R \right\rangle}, \label{SK}
\end{equation}
where $e^3$ is vierbein 3-form, and $\bar{\psi}_{L/R}$ are defined as
\begin{equation}
\bar{\psi}_{L/R} = \psi_{L/R}^{\dagger}\gamma_0 = -i\tilde{\psi}_{L/R}i\gamma_0= \mp\tilde{\psi}_{L/R}\gamma_0.
\end{equation}
Here outer products between differential forms are implicitly assumed.

One can write down the action for gravity as
\begin{align}
S_{Gravity} \sim \int{\left\langle ie^{2}(R + \frac{\Lambda}{24}e^2) \right\rangle}, 
\end{align}
where $e^2$ is vierbein 2-form, $R = d\omega + \omega^2$ is spin connection curvature 2-form, and  $\Lambda$ is cosmological constant.

The Yang-Mills action is written as
\begin{align}
&S_{Yang-Mills} =  S_{WL} + S_{WR} + S_{J} + S_{STR}, \label{SYM} \\
&S_{WL} \sim \int{\left\langle(e^2F_{WL})^2 \right\rangle / \left\langle ie^4   \right\rangle}, \\
&S_{WR} \sim \int{\left\langle(e^2F_{WR})^2 \right\rangle / \left\langle ie^4   \right\rangle}, \\
&S_{J} \sim \int{\left\langle(e^2F_{J})^2 \right\rangle / \left\langle ie^4 \right\rangle}, \\
&S_{STR} \sim \int{\left\langle(e^2F_{STR})^2 \right\rangle / \left\langle ie^4 \right\rangle}, 
\end{align}
where $e^4$ is vierbein 4-form. 

From an effective field theory point of view, an infinite number of terms allowed by symmetry requirements should be included in a generalized action. The gravity and Yang-Mills actions above are the first few order terms\cite{WL1} that are relevant in low-energy limit. 

The Clifford algebra elements, which are related to left-($e$, $\omega$, $W_L$, $W_R$) and right-($C$, $G$)sided gauge fields, are formally assigned to two sets of Clifford algebras in Yang-Mills action (and other actions without spinor fields). Elements from different sets formally commute with each other. Here $\left\langle \cdots\right\rangle$ means scalar part of both sets.

It's understood that 4-form factor $d^4x$ in one of $e^2F$ in each Yang-Mills term should be canceled out by 4-form factor $d^4x$ in the denominator before any further outer multiplication of differential forms as 
\begin{align}
&\int{\left\langle\frac{e^2F}{\left\langle ie^4\right\rangle}e^2F \right\rangle}. 
\end{align}
In this way, the Yang-Mills action is a diffeomorphism-invariant integration of 4-form on 4-dimensional space-time manifold. 

There is no explicit Hodge dual in Yang-Mills action. Vierbein plays the role of Hodge dual, when it acquires nonzero vacuum expectation value (VEV) in the case of flat space-time, which will be discussed in next section.

Yukawa and boson portions of the action will be subjects of later chapters.

\subsection{Local Lorentz Symmetry Breaking and Minkowskian Space-time}
Up to this point, the action is constructed in {\it curved} space-time, with space-time dependent vierbein and spin connection. In a vacuum with zero cosmological constant $\Lambda = 0$, vierbein field $e$ acquires a nonzero Minkowskian flat space-time VEV
\begin{align}
<0|e|0> = \delta_{\mu}^a \gamma_a dx^{\mu} = \gamma_{\mu} dx^{\mu},
\end{align}
while VEV of spin connection is zero
\begin{align}
<0|\omega|0> = 0.
\end{align}

The space-time metric reduces to
\begin{equation}
g_{\mu\nu} = \left\langle e_{\mu}e_{\nu} \right\rangle = \eta_{\mu\nu}.
\end{equation}

The soldering form $\gamma_{\mu} dx^{\mu}$ breaks the {\it independent} local Lorentz gauge invariance (on Clifford components with Roman indices such as in $\gamma_a\gamma_b$) and diffeomorphism invariance (on differential forms with Greek indices such as in $dx^{\mu}$). The action is left with a residual {\it global} Lorentz symmetry, with synchronized Clifford space and $x$ coordinate space global Lorentz rotations. Actually the specific VEV form $\gamma_{\mu} dx^{\mu}$ is a result of coordinating the above two kinds of global rotations. 

With the substitution of vierbein and spin connection with their VEVs, the spinor kinetic action(\ref{SK}) in flat Minkowskian space-time can be rewritten as
\begin{equation}
S_{Spinor-Kinetic} = \int {\left\langle \bar{\psi}_L \gamma^{\mu} D_{\mu} \psi_L
+ \bar{\psi}_R \gamma^{\mu} D_{\mu} \psi_R \right\rangle d^{4}x}, \label{ELEC}
\end{equation}
where
\begin{equation}
D_{\mu}\psi_{L/R} = (\partial_{\mu} + W_{L/R\mu})\psi_{L/R} + \psi_{L/R} (C_{\mu} - G_{\mu}).
\end{equation}

Similarly, the Yang-Mills action(\ref{SYM}) can be rewritten as
\begin{align}
S_{Yang-Mills} = 
&-\frac{1}{4g_{WL}^2}\int { F_{WL\mu\nu}^kF_{WL}^{\mu\nu k} d^{4}x} \\
&-\frac{1}{4g_{WR}^2}\int { F_{WR\mu\nu}F_{WR}^{\mu\nu} d^{4}x} \\
&-\frac{1}{4g_{J}^2}\int { F_{J\mu\nu}F_{J}^{\mu\nu} d^{4}x} \\
&-\frac{1}{4g_{STR}^2}\int { F_{STR\mu\nu}^kF_{STR}^{\mu\nu k} d^{4}x},
\end{align}
where $g_{WL}$, $g_{WR}$, $g_{J}$, and $g_{STR}$ are dimensionless gauge coupling constants. 

In the following chapters, however, we will stay with local Lorentz gauge invariant curved space-time formulation.

\subsection{Relation to Conventional Matrix Formulation}

A map can be constructed by placing the Dirac column spinor $\hat{\psi}$ in one-to-one correspondence with the algebraic spinor $\psi$. And the mappings for the operators are
\begin{align}
\hat{\gamma}^\mu\hat{\psi}  &\leftrightarrow  \gamma^\mu\psi, (\mu =0,1,2,3)\\
\hat{i}\hat{\psi}  &\leftrightarrow  \psi i, \\
\hat{\gamma}^5\hat{\psi}  &\leftrightarrow  -i\psi i
\end{align}
where $\hat{i}$ is the conventional unit imaginary number, and $\hat{\gamma}^\mu$ and $\hat{\gamma}^5$ are the Dirac matrix operators. 

We will not go into the details of further mappings \cite{WL1} in this paper.

\section{Majorana Bosons}

\subsection{Flavor Projection Operators}

With the purpose of studying 3 generations of fermions, we turn to another kind of Clifford algebra involving ternary communication relationships\cite{NA1, NA2} rather than the usual binary ones. Let's consider ternary $C\!\ell_{T1}$, which is defined by 
\begin{align}
&[\zeta, \zeta, \zeta] = \zeta^3 = 1,
\end{align}
with $\zeta$ commuting with $C\!\ell_{0,6}$
\begin{align}
&\zeta\gamma_j - \gamma_j\zeta = 0,\\
&\zeta\Gamma_j - \Gamma_j\zeta = 0.
\end{align}

Flavor projection operators are define by
\begin{align}
P_{1} &= \frac{1}{3}(1+e^{\sigma'+\sigma}\zeta+e^{-\sigma'-\sigma}\zeta^2) \\
			&= \frac{1}{3}P_l(1+\zeta+\zeta^2)
			+ \frac{1}{3}P_q(1+e^{-\sigma}\zeta+e^{\sigma}\zeta^2),  \label{IDEM7}\\
P_{2} &= \frac{1}{3}(1+e^{\sigma'}\zeta+e^{-\sigma'}\zeta^2) \\
			&= \frac{1}{3}P_l(1+e^{-\sigma}\zeta+e^{\sigma}\zeta^2)
			+ \frac{1}{3}P_q(1+e^{\sigma}\zeta+e^{-\sigma}\zeta^2),  \label{IDEM8}\\
P_{3} &= \frac{1}{3}(1+e^{\sigma'-\sigma}\zeta+e^{-\sigma'+\sigma}\zeta^2) \\
			&= \frac{1}{3}P_l(1+e^{\sigma}\zeta+e^{-\sigma}\zeta^2)
			+ \frac{1}{3}P_q(1+\zeta+\zeta^2),  \label{IDEM9}
\end{align}
where 
\begin{align}
&P_{1} + P_{2} + P_{3} = 1, \\
&P_{j}P_{k} = \delta_{jk}, \quad (j, k = 1, 2, 3), \\
&\sigma = \frac{2\pi}{3}i, \sigma' = \frac{2\pi}{3}i', \\
&i'=\frac{1}{2}(i+3J), i'^2=-1,
\end{align}
and $P_l$ and $P_q$ are lepton and quark projection operators, respectively. 

We label 3 generations of spinors as $\psi_{L/Rj}$. They are valued in $C\!\ell_{0,6}$. The spinor kinetic action involves 3 families of fermions as
\begin{align}
S_{Spinor-Kinetic} \sim &\int{\left\langle \bar{\psi}_{Lj}ie^3 D\psi_{Lj}P_{j} + \bar{\psi}_{Rj}ie^3 D\psi_{Rj}P_{j} \right\rangle}.
\end{align}
Here $\left\langle \cdots\right\rangle$ means scalar part of both $C\!\ell_{0,6}$ and $C\!\ell_{T1}$.
There is no flavor-mixing cross term in kinetic action. Flavor mixing is the subject of next section. It is induced via Majorana Boson fields.

It should be noted that there is another Clifford algebra based approach called spin-charge-family theory\cite{BORS}. It predicts a fourth family, coupled to the observed three families. The theory offers the explanation for the origin of the Higgs field and the Yukawa couplings. It also predicts a second group of four families, with the lowest of these four families explaining the origin of dark matter. 

\subsection{Majorana Yukawa Action}

Fields in Majorana boson section interact with right-handed fermions only.
The Lorentz, isospin, and color singlet Majorana boson section contains two fields
\begin{align}
&\phi_{MAJ} = \phi_{}^{\nu} + \Phi.
\end{align}
The neutrino Higgs field $\phi^{\nu}$ is valued in Clifford space spanned by 2 trivectors
\begin{align}
&\{\Gamma_0P_l, i\Gamma_0P_l\}.
\end{align}
It obeys gauge transformation rules
\begin{align}
&\phi_{}^{\nu}\quad\rightarrow\quad  e^{- \check{\Theta}_{WR} - \Theta_{B-L}}\phi_{}^{\nu} e^{\check{\Theta}_{WR} + \Theta_{B-L}}, 
\end{align}
where
\begin{align}
&\check{\Theta}_{WR} = \frac{1}{2}\epsilon_{WR} i
\end{align}
shares rotation angle $\epsilon_{WR}$ with
\begin{align}
&\Theta_{WR} = \frac{1}{2}\epsilon_{WR}\Gamma_1\Gamma_2. 
\end{align}

Boson field 
\begin{align}
\Phi = \Phi_{12} + \Phi_{13A} + \Phi_{13B} + \Phi_{22} + \Phi_{23}
\end{align}
is valued in Clifford space spanned by scalar and pseudoscalar
\begin{align}
&\{ 1, i\}.
\end{align}
It is invariant under all local gauge transformations.

We can write Majorana Yukawa action of right-handed fermions as
\begin{align}
S_{Majorana-Yukawa} \quad \sim \quad
&y_{11}\int{\left\langle \phi_{}^{\nu}P_{1} \bar{\nu}_{R1}e^4e^{\varepsilon_{11}\Gamma_1\Gamma_2}\Gamma_2\Gamma_3\nu_{R1}P_{1} \right\rangle} \\
+ &y_{23}\int{\left\langle \phi_{}^{\nu}P_{2} \bar{\nu}_{R2}e^4e^{\varepsilon_{23}\Gamma_1\Gamma_2}\Gamma_2\Gamma_3\nu_{R3}P_{3} \right\rangle} + h.c.  \label{MIX}\\
+ &Y_{12}\int{\left\langle \Phi_{12} P_{1}\bar{u}_{R1}e^4d_{R2}P_{2}\bar{u}_{R2}e^4d_{R1} \right\rangle/\left\langle ie^4 \right\rangle} + h.c.\\ 
+ &Y_{13A}\int{\left\langle \Phi_{13A} P_{1}\bar{u}_{R1}e^4e_{R3}P_{3}\bar{\nu}_{R3}e^4d_{R1} \right\rangle/\left\langle ie^4 \right\rangle} + h.c.\\ 
+ &Y_{13B}\int{\left\langle \Phi_{13B} P_{1}\bar{\nu}_{R1}e^4d_{R3}P_{3}\bar{u}_{R3}e^4e_{R1} \right\rangle/\left\langle ie^4 \right\rangle} + h.c.\\ 
+ &Y_{22}\int{\left\langle \Phi_{22} P_{2}\bar{u}_{R2}e^4e_{R2}P_{2}\bar{\nu}_{R2}e^4d_{R2} \right\rangle/\left\langle ie^4 \right\rangle} + h.c.\\ 
+ &Y_{23}\int{\left\langle \Phi_{23} P_{2}\bar{\nu}_{R2}e^4e_{R3}P_{3}\bar{\nu}_{R3}e^4e_{R2} \right\rangle/\left\langle ie^4 \right\rangle} + h.c.,
\end{align}
where $y_{jk}$ and $Y_{jk}$ are Majorana Yukawa coupling constants, and $e^{\varepsilon_{jk}\Gamma_1\Gamma_2}$ are phase factors. 

There are four fermions in the Yukawa terms of $\Phi_{jk}$, while Higgs boson $\phi_{}^{\nu}$ interacts with two fermions. The four-fermion Yukawa coupling constants $Y_{jk}$ are of mass dimension $-3$. Thus four-fermion Yukawa terms are nonrenormalizable. A later section will discuss the effective theory point of view and the issue of nonrenormalizability. 

Since $\frac{2\pi}{3}i$ phases in flavor projection operators anticommute with Clifford-odd fields, there are properties
\begin{align}
&P_{1}\phi_{}^{\nu} = \phi_{}^{\nu}P_{1}, \\
&P_{2}\phi_{}^{\nu} = \phi_{}^{\nu}P_{3}, \\
&P_{3}\phi_{}^{\nu} = \phi_{}^{\nu}P_{2},
\end{align}
according to the definition of flavor projection operators (\ref{IDEM7}, \ref{IDEM8}, \ref{IDEM9}). Therefore, there are flavor-mixing terms (\ref{MIX}) between 2nd and 3rd generation neutrinos, as evidenced in the observation of neutrino oscillations\cite{FUK, AHM, EGU}. 

Likewise, allowable four-fermion Yukawa terms are also dictated by the properties of flavor projection operators. Additionally, we require that the four-fermion combination should accommodate a global $U(1)$ symmetry, which will be studies in next section. The five $\Phi_{12}, \Phi_{13A}, \Phi_{13B}, \Phi_{22},$ and $\Phi_{23}$ Yukawa terms, with fermion configurations 
\begin{align}
&\bar{u}_Rs_R\bar{c}_Rd_R \quad\rightarrow\quad \Phi_{12}, \\
&\bar{u}_R\tau_R\bar{\nu}_{\tau R}d_R \quad\rightarrow\quad \Phi_{13A}, \\
&\bar{\nu}_{e R}b_R\bar{t}_{R}e_R \quad\rightarrow\quad \Phi_{13B}, \\
&\bar{c}_R\mu_R\bar{\nu}_{\mu R}s_R, \quad\rightarrow\quad \Phi_{22}, \\
&\bar{\nu}_{\mu R}\tau_R\bar{\nu}_{\tau R}\mu_R \quad\rightarrow\quad \Phi_{23},
\end{align}
are the ones satisfying both conditions. 

After $\phi_{}^{\nu}$ and $\Phi$ acquire nonzero VEVs, which will be investigated in later section, the flavor mixing between right-handed fermions is represented by neutrino Majorana mass terms and four-fermion interaction terms. Higher order processes can introduce further effective mixing between generations. One may potentially couple above effects with appropriate choices of Majorana and electroweak Yukawa coupling constants to explain the quite different patterns of CKM and PMNS\footnote{See ref. \cite{TRI1, TRI2} for attempts to explain the PMNS matrix pattern.} matrices.

\subsection{Flavor-Related Global $U(1)$ Symmetry}
As mentioned earlier, $\Phi$ boson is invariant under all gauge transformations related to gauge interactions. Nevertheless, for $\Phi_{12}$ there is a flavor-related {\it global} $U(1)$ symmetry under the following transformations
\begin{align}
&\Phi_{12} \quad\rightarrow\quad   \Phi_{12} e^{\theta_{12} i}, \\
&u_{1} = u_{}  \quad\rightarrow\quad  u_{}e^{\theta_u i}, \\
&d_{1} = d_{}  \quad\rightarrow\quad  d_{}e^{\theta_d i}, \\
&u_{2} = c_{}  \quad\rightarrow\quad  c_{}e^{\theta_c i}, \\
&d_{2} = s_{}  \quad\rightarrow\quad  s_{}e^{\theta_s i}, 
\end{align}
where
\begin{align}
&\theta_{12} = (\theta_u - \theta_d) - (\theta_c - \theta_s),
\end{align}
and $u, d, c,$ and $s$ are up, down, charm, and strange quarks. The phase $\theta_{12}$ measures rotation angle difference between first and second generation quarks in $\bar{u}_Rs_R\bar{c}_Rd_R$. 

Because of 
\begin{align}
&\bar{u}_R = u^{\dagger}_R\gamma_0, \\
&\bar{c}_R = u^{\dagger}_R\gamma_0,
\end{align}
moving a phase factor $\theta i$ around Clifford-odd $\gamma_0$ changes its sign as
\begin{align}
&(\theta i)u^{\dagger}_R\gamma_0 = u^{\dagger}_R\gamma_0(-\theta i), \\ 
&(\theta i)c^{\dagger}_R\gamma_0 = c^{\dagger}_R\gamma_0(-\theta i). \\ 
\end{align}
This is the reason of the specific sign for each $\theta$ in $\theta_{12}$.

By the same token, $\bar{u}_R\tau_R\bar{\nu}_{\tau R}d_R, \bar{\nu}_{e R}b_R\bar{t}_{R}e_R, \bar{c}_R\mu_R\bar{\nu}_{\mu R}s_R,$ and $\bar{\nu}_{\mu R}\tau_R\bar{\nu}_{\tau R}\mu_R$ correspond to phases
\begin{align}
&\theta_{13A} = (\theta_u - \theta_d) - (\theta_{\nu_\tau} - \theta_\tau), \\
&\theta_{13B} = (\theta_{\nu_e} - \theta_e) - (\theta_t - \theta_b), \\
&\theta_{22} = (\theta_c - \theta_s) - (\theta_{\nu_{\mu}} - \theta_{\mu}), \\
&\theta_{23} = (\theta_{\nu_\mu} - \theta_\mu) - (\theta_{\nu_\tau} - \theta_\tau).
\end{align}

In the event of spontaneous symmetry breaking (SSB), there will be a massive sigma mode and a massless Nambu-Goldstone mode for each $\Phi_{ij}$. As opposed to the Higgs mechanism, the Nambu-Goldstone mode is not 'eaten' by gauge field. 

Notice that above global symmetry is an approximate symmetry. The electroweak section spoils the symmetries $\theta_{12}, \theta_{13A}, \theta_{13B}, \theta_{22},$ and $\theta_{23}$ explicitly\footnote{If we hold $\theta_{\nu_e}$, $\theta_{\nu_\mu}$, and $\theta_{\nu_\tau}$ equal to zero, neutrino Higgs field $\phi^{\nu}$ Yukawa terms do not explicitly break the global symmetries.}.  The Nambu-Goldstone modes are not exactly massless. The sizes of the masses grow with the strength of the explicit symmetry breaking. A not-quite-massless would-be Nambu-Goldstone particle for an approximate symmetry is often called a pseudo-Nambu-Goldstone boson (pNGB). These flavor-related pNGBs represent phase differences between four different fermions.

Since boson fields $\Phi_{ij}$ are free from direct gauge interactions, they are potential dark matter candidates. $\Phi_{23}$ might be the prime candidate due to large Majorana masses of two neutrinos ($\nu_{\mu R}$ and $\nu_{\tau R}$) and the resulted suppressed tree-level decay rate.

\subsection{Symmetry Breaking and Majorana Masses}

The Majorana Boson action reads\footnote{We write down the action as if $\Phi = \Phi_{12}$. The treatment of other bosonic fields $\Phi_{13A}, \Phi_{13B}, \Phi_{22},$ and $\Phi_{23}$ should follow the same logic as $\Phi_{12}$.}
\begin{align}
S_{Majorana-Bosons} =  S_{Majorana-Kenetic} -  V_{Majorana},
\end{align}
with
\begin{align}
&S_{Majorana-Kenetic}(\phi^{\nu}) \sim  
\int{\left\langle(e^3D\phi^{\nu \dagger})(e^3D\phi^{\nu}) \right\rangle 
/ \left\langle ie^4  \right\rangle }, \\
&V_{Majorana-Bosons}(\phi^{\nu}, -\mu_{\nu}^2, \lambda_{\nu}) \sim  
\int{ (-\mu_{\nu}^2 |\phi^{\nu}|^2 + \lambda_{\nu} |\phi^{\nu}|^4) \left\langle ie^4\right\rangle},
\end{align}
and 
\begin{align}
&S_{Majorana-Kenetic}(\Phi) \sim  
\int{\left\langle(e^3D\Phi^{\dagger})(e^3D\Phi)\right\rangle 
/ \left\langle ie^4  \right\rangle }, \\
&V_{Majorana-Bosons}(\Phi, -\mu_{\Phi}^2, \lambda_{\Phi}) \sim  
\int{ (-\mu_{\Phi}^2 |\Phi|^2 + \lambda_{\Phi} |\Phi|^4) \left\langle ie^4\right\rangle},
\end{align}
where
\begin{align}
&D\phi^{\nu} = (d - \check{W}_R - C)\phi^{\nu} + \phi^{\nu}(\check{W}_R + C), \label{MJNU} \\
&D\Phi = d\Phi, \\
&\check{W}_R =  \check{W}_{R\mu}dx^{\mu}= \frac{1}{2}W^3_{R\mu}i dx^{\mu}. 
\end{align}
Notice that $\phi^{\nu}$ and $\Phi$ have negative $-\mu_{\nu}^2$ and $-\mu_{\Phi}^2$. It means that $\phi^{\nu}$ and $\Phi$ acquire nonzero VEVs as
\begin{align}
&<0|\phi^{\nu}|0> = \frac{1}{\sqrt{2}}\upsilon_{\nu}\Gamma_0P_le^{\alpha i} = \frac{1}{\sqrt{2}}\frac{\mu_{\nu}}{\sqrt{\lambda_{\nu}}}\Gamma_0P_le^{\alpha i}, \\
&<0|\Phi|0> = \frac{1}{\sqrt{2}}\upsilon_{\Phi}e^{\alpha_{12} i} = \frac{1}{\sqrt{2}}\frac{\mu_{\Phi}}{\sqrt{\lambda_{\Phi}}}e^{\alpha_{12} i},
\end{align}
As a result, the gauge symmetry related to gauge field 
\begin{align}
&Z'_{\mu} = W^3_{R\mu} - C^J_{\mu}, 
\end{align}
and the global symmetry of $\Phi$ are spontaneously broken. Notice that the minus sign in above equation stems from the fact that $JP_l = -iP_l$. The symmetries of $U(1)_{WR} \otimes U(1)_{B-L}$ are reduced to the residual hypercharge $U(1)_Y$ symmetry with transformations
\begin{align}
&\psi_{L}\rightarrow \psi_{L} e^{\frac{1}{2}\epsilon_Y J}, \\
&\psi_{R}\rightarrow e^{\frac{1}{2}\epsilon_Y\Gamma_1\Gamma_2}\psi_{R} e^{\frac{1}{2}\epsilon_Y J},
\end{align}
where a shared rotation angle $\epsilon_Y$ synchronizes the gauge transformations. 

After replacement of $\phi^{\nu}$ and $\Phi$ with their VEVs, the Majorana Yukawa action reduces to 
\begin{align}
S_{Majorana-Yukawa} \quad \sim \quad
&\int{\left\langle m_{11}P_{1}\bar{\nu}_{R1}e^4\Gamma_2\Gamma_3\nu_{R1}P_{1}\Gamma_0P_l \right\rangle}, \\
+ &\int{\left\langle m_{23}P_{2}\bar{\nu}_{R2}e^4\Gamma_2\Gamma_3\nu_{R3}P_{3}\Gamma_0P_l \right\rangle} + h.c., \\
+ &\frac{1}{\sqrt{2}}Y_{12}\upsilon_{\Phi}\int{\left\langle e^{\alpha_{12} i}P_{1}\bar{u}_{R1}e^4d_{R2}P_{2}\bar{u}_{R2}e^4d_{R1} \right\rangle/\left\langle ie^4 \right\rangle} + h.c.,
\end{align}
with Majorana masses
\begin{align}
&m_{11} = \frac{1}{\sqrt{2}}y_{11}\upsilon_{\nu}e^{(\alpha + \varepsilon_{11})i}, \\
&m_{23} = \frac{1}{\sqrt{2}}y_{23}\upsilon_{\nu}e^{(\alpha + \varepsilon_{23})i}.
\end{align}

Neutrino Majorana masses are much heavier than neutrino Dirac masses, if we assume
\begin{align}
&y_{jk}\upsilon_{\nu} >> y^{\nu} \upsilon,
\end{align}
where constants $y^{\nu}$ and $\upsilon$ are electroweak Higgs counterparts, which will be defined in later section. Because of the hierarchy, very small effective masses are generated for neutrinos, known as seesaw mechanism. 

Now we express gauge fields $W^3_{R}$ and $C^J_{}$ in terms of $B_{}$ and $Z'_{}$ \footnote{Since the gauge fields here are defined to absorb gauge coupling constants, the formulas here look a bit different from the conventional ones. But it is only a notational difference.}
\begin{align}
&W^3_{R\mu} = B_{\mu} + (cos\theta_W')^2 Z'_{\mu}, \\
&C^J_{\mu} = B_{\mu} - (sin\theta_W')^2 Z'_{\mu}, 
\end{align}
where
\begin{align}
&cos\theta_W' = \frac{g_{WR}}{g_{Z'}}, \\
&sin\theta_W' = \frac{g_{J}}{g_{Z'}}, \\
&g_{Z'} = \sqrt{g_{WR}^2 + g_{J}^2}.
\end{align}

Gauge field $B$ remains massless with an effective coupling of 
\begin{align}
&g_{B} = \frac{g_{WR}g_{J}}{g_{Z'}},
\end{align}
while gauge field $Z'$ acquires a mass from neutrino part of the Majorana Kinetic action
\begin{align}
&M_{Z'} = \frac{1}{2}\upsilon_{\nu}g_{Z'}.
\end{align}
Higgs boson $\phi^{\nu}$ and the sigma mode of $\Phi$ acquire masses
\begin{align}
&m_{h\nu} = \sqrt{2}\mu_{\nu}, \\
&m_{\Phi} = \sqrt{2}\mu_{\Phi}.
\end{align}

If we assume that $\upsilon_{\nu} >> \upsilon$ and $\upsilon_{\Phi} >> \upsilon, $ gauge boson $Z',$ Higgs boson $\phi^{\nu},$ and sigma mode of $\Phi$ would be too heavy to be detected at electroweak energy scale. On the other hand, the pNGB of $\Phi$ is not exactly massless, since the electroweak sector explicitly breaks the global symmetry and can generate mass for it. The size of the $\Phi$ pNGB mass is proportional to electroweak scale. Hence it is detectable at LHC. 

The experiments at LHC recently indicated a diphoton resonance at about 750 Gev\cite{H750A, H750C}, in addition to the earlier finding of Higgs boson with $m_h = 125$ Gev\cite{H125A, H125C}. Various explanations have been offered\cite{H750AAA1, H750AAA2, H750AA, H750AAA6, H750BB, H750BB2, H750BB3, H750AAA4, H750CC, H750DD, H750EE, H750EE2, H750AAA5, H750FF, H750AAA3, H750AAA7}. Scenarios with either an isospin singlet state or an isospin doublet state usually need an extended particle content to accommodate the observed signal.

The LHC 750 GeV diphoton resonance might possibly be identified as the $\Phi_{13B}$ ($\Phi_{13A}$ or $\Phi_{22}$) pNGB with mass around 750 Gev. It results from SSB of the flavor-related global $U(1)$ symmetry involving leptons and quarks. 

A resonance starts with four quarks (two quark-antiquark pairs) produced by two gluons. Two of the quarks interact with VEV $<0|\Phi_{13B}^\dagger|0>$ ($<0|\Phi_{13A}^\dagger|0>$ or $<0|\Phi_{22}^\dagger|0>$)and turn into two leptons. And then the four fermions turn into a $\Phi_{13B}$ ($\Phi_{13A}$ or $\Phi_{22}$) pNGB via the four-fermion Yukawa term. The pNGB propagates and finally decays in a reverted generation process. The reverted process has two external photon lines, instead of two external gluon lines.  

The diphoton decay of the pseudo-Nambu-Goldstone boson is loop induced, since tree-level decay is suppressed by large Majorana mass of the right-handed neutrino $\nu_{e R}$ ($\nu_{\tau R}$ or $\nu_{\mu R}$). The width of the 750 Gev resonance might be due to the combined resonance signals coming from $\Phi_{13B}$, $\Phi_{13A}$, and $\Phi_{22}$ pNGBs with different resonance masses. 

\subsection{Four-Fermion Condensation}
Boson sectors might be just an effective Ginzburg-Landau-type description of the low energy physics represented by composite boson  fields. One approach is to assume effective four-quark interactions strong enough to induce top quark-antiquark condensation into composite electroweak Higgs fields\cite{TOP0, TOP1, TOP2, TOP3}, via dynamical symmetry breaking mechanism in Nambu-Jona-Lasinio\cite{NJL} (NJL) like models.

The four-quark contact term in top condensation model is 
\begin{align}
&\int{\left\langle \bar{q}_L e q_L \bar{t}_R e^3 t_R \right\rangle}, 
\end{align}
where $q_L = t_L + b_L$, $e$ and $e^3$ are vierbein 1-form and 3-form.

Likewise, the Majorana boson fields are also collective excitations of underlying composite spinors. For example, $\phi^{\nu}$ and $\Phi_{12}$ are effective representation of two-neutrino and four-quark condensations
\begin{align}
\check{\phi}^{\nu} &= y_{11}P_{1}\bar{\nu}_{R1}e^4\Gamma_2\Gamma_3\nu_{R1}P_{1}, \\
& + y_{23}P_{2}\bar{\nu}_{R2}e^4\Gamma_2\Gamma_3\nu_{R3}P_{3} + h.c., \\
\check{\Phi_{12}} &= Y_{12}P_{1}\bar{u}_{R}e^4s_{R}P_{2}\bar{c}_{R}e^4d_{R}/\left\langle ie^4 \right\rangle + h.c..
\end{align}

The four-neutrino and eight-quark interactions are
\begin{align}
&\int{\left\langle \check{\phi}^{\nu \dagger} \check{\phi}^{\nu} \right\rangle / \left\langle ie^4 \right\rangle}
+\int{\left\langle \check{\Phi_{12}}^{\dagger}\check{\Phi_{12}}  \right\rangle / \left\langle ie^4 \right\rangle}.
\end{align}

A collective mode of Higgs field $\phi^{\nu}$ is determined as the pole of bosonic channel of the four-neutrino interaction by summing to infinite order chains of bubble perturbation diagrams. The leading order calculation goes by different names such as random-phase approximation, Bethe-Salpeter T-matrix equation, and 1/N expansion.

If the Majorana bosonic fields $\Phi_{ij}$ are indeed collective excitations of the underlying four fermions, the first order approximation would involve summing to infinite order chains of 'bubble' diagrams, linked together via eight-fermion contact interactions. Each 'bubble' contains four lines of fermion propagators, rather than two lines.

Condensations $\bar{u}_Rs_R\bar{c}_Rd_R, \bar{u}_R\tau_R\bar{\nu}_{\tau R}d_R, \bar{\nu}_{e R}b_R\bar{t}_{R}e_R, \bar{c}_R\mu_R\bar{\nu}_{\mu R}s_R,$ and $\bar{\nu}_{\mu R}\tau_R\bar{\nu}_{\tau R}\mu_R$ are potential dark matter candidates, since they are free from direct gauge interactions. The prime candidate might be the four-lepton condensation $\bar{\nu}_{\mu R}\tau_R\bar{\nu}_{\tau R}\mu_R$, due to large Majorana masses of two neutrinos ($\nu_{\mu R}$ and $\nu_{\tau R}$) and the resulted suppressed tree-level decay rate.  

\subsection{The Issue of Nonrenormalizability}
The four-fermion Yukawa terms of $\Phi$ and four/eight-quark contact interactions are nonrenormalizable in the conventional sense.  

For nonrenormalizable models, the renomalization procedure can be made only at the cost of adding increasing numbers of term to the original Lagrangian. In principle, there is no problem with a theory having an infinite number of coupling constants as an effective field theory\cite{WEIN}. However, the NJL model is often regarded as regularization-dependent and its predictability is called into question. 

A novel strategy for handling divergences is called implicit regularization \cite{IMPL}. It avoids the critical step of explicit evaluation of divergent integrals. The finite parts are separated from the divergent ones and integrated free from effects of regulation. The application to NJL model reveals that it can be ambiguity-free and symmetry-preserving can be obtained, making the NJL model predictive.

Likewise, we expect that models with four-fermion Yukawa interactions or eight-fermion contact terms are as predictive as renormalizable theories. 

\section{Electroweak Bosons}

\subsection{Electroweak Bosons and Yukawa Action}
Electroweak boson field $\phi_{EW}$ interacts with both left-handed and right-handed fermions,  while Majorana boson field $\phi_{MAJ}$ interacts with right-handed fermions only. Electroweak Boson field $\phi_{EW}$ spans the whole 32 component $C\!\ell_{0,6}$ even space. It obeys gauge transformation rules
\begin{align}
&\phi_{EW}\quad\rightarrow\quad  e^{\Theta_{LOR} + \Theta_{WL}}\phi_{EW} e^{-\Theta_{LOR} - \Theta_{WR}}.
\end{align}

It can be broken down into three fields as
\begin{align}
&\phi_{EW} = \phi_{S} + \phi_{P} + \phi_{AT}, 
\end{align}
with scalar $\phi_{S}$ valued in Clifford space spanned by 4 multivectors
\begin{align}
&\{1, \Gamma_j\Gamma_k; \quad j, k = 1, 2, 3, j\neq k \}, 
\end{align}
pseudoscalar $\phi_{P}$ valued in Clifford space spanned by 4 multivectors
\begin{align}
&\{i, i\Gamma_j\Gamma_k; \quad j, k = 1, 2, 3, j\neq k\}, 
\end{align}
and antisymmetric tensor $\phi_{AT}$ valued in Clifford space spanned by $4*6 = 24$ multivectors
\begin{align}
&\{\gamma_a\gamma_b, \gamma_a\gamma_b\Gamma_j\Gamma_k; 
\quad j, k = 1, 2, 3, j\neq k, 
a, b = 0, 1, 2, 3, a\neq b\}.
\end{align}

The scalar and pseudoscalar electroweak Higgs fields $\phi_{S}$ and $\phi_{P}$ transform as 
\begin{align}
&\phi_{S/P}\quad\rightarrow\quad  e^{\Theta_{WL}}\phi_{S/P} e^{- \Theta_{WR}},
\end{align}
while the antisymmetric tensor electroweak boson field $\phi_{AT}$ transforms as 
\begin{align}
&\phi_{AT}\quad\rightarrow\quad  e^{\Theta_{LOR} + \Theta_{WL}}\phi_{AT} e^{-\Theta_{LOR} - \Theta_{WR}}.
\end{align}
Notice that $\phi_{AT}$ is {\it not a Lorentz singlet}, since it's not invariant under local Lorentz gauge transformations.

We can write electroweak Yukawa action of fermions as
\begin{align}
&S_{Electroweak-Yukawa} \quad \sim \quad \\
&\int{\left\langle \bar{\psi}_{Lj}ie^4 \phi_{EW}(y^{\nu}_{j}\nu_{Rj} + y^{e}_{j}e_{Rj}
+ y^{u}_{j}u_{Rj} + y^{d}_{j}d_{Rj}
)i P_{j} \right\rangle} \\
&+\int{\left\langle (y^{\nu}_{j}\bar{\nu}_{Rj} + y^{e}_{j}\bar{e}_{Rj}
+ y^{u}_{j}\bar{u}_{Rj} + y^{d}_{j}\bar{d}_{Rj}
)ie^4 \bar{\phi}_{EW}\psi_{Lj}i P_{j} \right\rangle},
\end{align}
where
\begin{align}
& \bar{\phi}_{EW}  = \gamma_0\phi_{EW}^{\dagger}\gamma_0 = \gamma_0\tilde{\phi}_{EW}\gamma_0,
\end{align}
and $y_{j}^{\nu}, y_{j}^{e}, y_{j}^{u},$ and $y_{j}^{d}$ are electroweak Yukawa coupling constants. 

\subsection{Electroweak Boson Action, Symmetry Breaking, and Dirac Mass}

Electroweak boson action reads
\begin{align}
S_{Electroweak-Bosons} =  S_{Electroweak-Kenetic} -  V_{Electroweak},
\end{align}
with
\begin{align} 
&S_{Electroweak-Kenetic}(\phi_{S}) \sim  
\int{\left\langle(e^3(\bar{D\phi_{S}}))(e^3D\phi_{S}) \right\rangle 
/ \left\langle ie^4  \right\rangle }, \\
&V_{Electroweak-Bosons}(\phi_{S}, -\mu_{S}^2, \lambda_{S}) \sim  
\int{ (-\mu_{S}^2 |\phi_{S}|^2 + \lambda_{S} |\phi_{S}|^4)\left\langle ie^4\right\rangle},
\end{align}
and 
\begin{align}
&S_{Electroweak-Kenetic}(\phi_{P}), 
V_{Electroweak}(\phi_{P}, -\mu_{P}^2, \lambda_{P}), \\
&S_{Electroweak-Kenetic}(\phi_{AT}), \\
&V_{Electroweak}(\phi_{AT}, +\mu_{AT}^2, \lambda_{AT}) \sim  
\int{ (\mu_{AT}^2 \left\langle\bar{\phi}_{AT}\phi_{AT}\right\rangle + \lambda_{AT} (\left\langle\bar{\phi}_{AT}\phi_{AT}\right\rangle)^2) \left\langle ie^4 \right\rangle},
\end{align}
where
\begin{align}
&D\phi_{P/S} = (d + W_L)\phi_{P/S} - \phi_{P/S}(W_R), \label{EWSP} \\
&D\phi_{AT} = (d + \omega + W_L)\phi_{AT} - \phi_{AT}(\omega + W_R), \label{EWAT} 
\end{align}
Notice that $\phi_{S}$ and $\phi_{P}$ have negative $-\mu_{S}^2$ and $-\mu_{P}^2$. It means that $\phi_{S}$ and $\phi_{P}$ acquire nonzero VEVs via SSB
\begin{align}
&<0|\phi_{S}|0> = \frac{1}{\sqrt{2}}\upsilon_S = \frac{1}{\sqrt{2}}\frac{\mu_{S}}{\sqrt{\lambda_{S}}}, \\
&<0|\phi_{P}|0> = \frac{1}{\sqrt{2}}\upsilon_Pi = \frac{1}{\sqrt{2}}\frac{\mu_{P}}{\sqrt{\lambda_{P}}}i.
\end{align}
The situation of $\phi_{AT}$ is a bit complicated, and will be discussed in later section. Let's for the moment assume that its VEV is zero.

After replacing $\phi_{S}, \phi_{P}, $ and $\phi_{AT}$ with their VEVs, the electroweak Yukawa action reduces to
\begin{align}
& \int{\left\langle  (\bar{\nu}_{j}ie^4m^{\nu}_{j}\nu_{j}i + \bar{e}_{j}ie^4m^{e}_{j}e_{j}i
+ \bar{u}_{j}ie^4m^{u}_{j}u_{j}i + \bar{d}_{j}ie^4m^{d}_{j}d_{j}i
) P_{j} \right\rangle} , 
\end{align}
where 'complex' (scalar plus pseudoscalar) Dirac masses are
\begin{align}
&m^{\nu/e/u/d}_{j} = \frac{1}{\sqrt{2}}y^{\nu/e/u/d}_{j}(\upsilon_S + \upsilon_Pi) = \frac{1}{\sqrt{2}}y^{\nu/e/u/d}_{j}\upsilon e^{\beta i},
\end{align}
with
\begin{align}
&\upsilon = \sqrt{\upsilon_S^2 + \upsilon_P^2}, \\
&tan(\beta) = \frac{\upsilon_P}{\upsilon_S}.
\end{align}
However the $e^{\beta i}$ phase factor can be canceled out via a global rotation of spinor
\begin{align}
&\psi \quad\rightarrow\quad e^{-\frac{1}{2}\beta i}\psi,
\end{align}
so that the fermion Dirac masses are 'real' (scalar) -valued. If scalar and pseudoscalar Higgs fields have different configurations of Yukawa coupling constants, the rotation angles are spinor ($\nu/e/u/d$) specific. 

Since the experiments at LHC indicated only one Higgs boson with $m_h = 125$ Gev\cite{H125A, H125C}, there could be two scenarios. Case one is that both scalar and pseudoscalar Higgs fields contribute to the electroweak symmetry breaking and their masses are degenerate
\begin{align}
m_h = m_S = m_P.
\end{align}
Case two is that only one of them acquires a nonzero VEV (with negative $-\mu^2$), which is the $m_h = 125$ Gev Higgs. The other maintains a zero VEV (with positive $\mu^2$), which is still waiting to be detected at LHC. 

Now we express gauge fields $W^3_{L},$ $B_{},$ and $W^3_{R}$ in terms of $A_{},$ $Z_{},$ and $Z'$ 
\begin{align}
&W^3_{L\mu} = A_{\mu} + (cos\theta_W)^2 Z_{\mu}, \\
&B_{\mu} = A_{\mu} - (sin\theta_W)^2 Z_{\mu}, \\
&W^3_{R\mu} = B_{\mu} + (cos\theta_W')^2 Z'_{\mu} = A_{\mu} - (sin\theta_W)^2 Z_{\mu} + (cos\theta_W')^2 Z'_{\mu},
\end{align}
where
\begin{align}
&cos\theta_W = \frac{g_{WL}}{g_{Z}}, \\
&sin\theta_W = \frac{g_{B}}{g_{Z}}, \\
&g_{Z} = \sqrt{g_{WL}^2 + g_{B}^2}.
\end{align}

Electromagnetic field $A$ remains massless with an effective coupling of 
\begin{align}
&g = \frac{g_{WL}g_{B}}{g_{Z}} = \frac{g_{WL}g_{WR}g_{J}}
{\sqrt{g_{WL}g_{WR} + g_{WL}g_{J} + g_{WR}g_{J}}},
\end{align}
while gauge field $Z$ acquires a mass 
\begin{align}
&M_{Z} = \frac{1}{2}\upsilon g_{Z}.
\end{align}

\subsection{Antisymmetric Tensor Boson}
The antisymmetric tensor boson is a bridge between gravity and electroweak sector. The strong connection between gravity field and electroweak Higgs field has also been studied in a geometrical 5D unification approach\cite{CAPOZ2}, which deduces all the known interactions from an induced symmetry breaking of the non-unitary GL(4)-group of diffeomorphisms. 

As stated earlier, the antisymmetric tensor field $\phi_{AT}$ is not invariant under Lorentz gauge transformations. Hence, its boson potential should involve Lorentz invariant
\begin{align}
\left\langle\bar{\phi}_{AT}\phi_{AT}\right\rangle = \left\langle \gamma_0\phi_{AT}^{\dagger}\gamma_0\phi_{AT}\right\rangle, 
\end{align}
as opposed to 
\begin{align}
|\phi_{AT}|^2 = \left\langle \phi_{AT}^{\dagger} \phi_{AT} \right\rangle,
\end{align}
which is not Lorentz invariant.

It's easy to see that $\left\langle\bar{\phi}_{AT}\phi_{AT}\right\rangle$ is not a positive definite quantity. Components of
\begin{align}
&\{\gamma_a\gamma_b, \gamma_a\gamma_b\Gamma_j\Gamma_k; 
\quad j, k = 1, 2, 3, j\neq k, 
a, b = 1, 2, 3, a\neq b\}, 
\end{align}
have positive 'metric' and components of
\begin{align}
&\{i\gamma_a\gamma_b, i\gamma_a\gamma_b\Gamma_j\Gamma_k; 
\quad j, k = 1, 2, 3, j\neq k, 
a, b = 1, 2, 3, a\neq b\},  
\end{align}
have negative 'metric'. 

We can divide $\phi_{AT}$ as
\begin{align}
&\phi_{AT} =  \phi_{ATs} + \phi_{ATp},
\end{align}
where $\phi_{ATs}$ and $\phi_{ATp}$ are valued in positive and negative 'metric' components, respectively. Thus we have
\begin{align}
&\left\langle\bar{\phi}_{AT}\phi_{AT}\right\rangle =  |\phi_{ATs}|^2 - |\phi_{ATp}|^2.
\end{align}

A zero VEV $<0|\phi_{AT}|0>$ is allowed only if $\mu_{AT}^2= 0$. On the other hand, 
nonzero VEV can be acquired for any value of $\mu_{AT}^2$, including $\mu_{AT}^2= 0$. Nonzero VEV simultaneously breaks electroweak and Lorentz symmetries. 

In the case of $\mu_{AT}^2= 0$, the four-boson-field term in the boson potential enforces
\begin{align}
&(\left\langle\bar{\phi}_{AT}\phi_{AT}\right\rangle)^2 =  (|\phi_{ATs}|^2 - |\phi_{ATp}|^2)^2 = 0.
\end{align}
Therefore, the VEV should be on the 'light cone', which means 
\begin{align}
&|\phi_{ATs}|^2 = |\phi_{ATp}|^2.
\end{align}

Replacing $\phi_{AT}$ with nonzero $<0|\phi_{AT}|0>$ in the boson kinetic action, we have a Lorentz symmetry breaking term
\begin{align} 
&\int{\left\langle(e^3(\omega<0|\phi_{AT}|0> - <0|\phi_{AT}|0>\omega) {\bar{}} )(e^3(\omega<0|\phi_{AT}|0> - <0|\phi_{AT}|0>\omega)) \right\rangle 
/ \left\langle ie^4  \right\rangle }.
\end{align}
This spin connection $\omega$ related term can contribute to space-time torsion equation. We call it 'dark spin current'. It is a counterpart of dark energy, with the former affecting space-time torsion and the later affecting space-time curvature.

Since we know that torsion could have gravitational and cosmological consequences\cite{CAPOZ, MODC}, it's worth further research on the above antisymmetric-tensor-induced scenario.

\section{Possible Grand Unification Symmetries}

We now explore more symmetries allowed by an algebraic spinor. Let's begin with general gauge transformations
\begin{equation}
\psi\quad\rightarrow\quad e^{\Theta}\psi e^{\Theta'},
\end{equation}
where $e^{\Theta}$ and $e^{\Theta'} \in C\!\ell_{0,6}$ are independent gauge transformations. Spinor bilinear
\begin{equation}
\left\langle \tilde{\psi}\gamma_0\psi \right\rangle \label{BILI}
\end{equation}
is invariant if
\begin{align}
&e^{\tilde{\Theta}}\gamma_0 e^{\Theta} = \gamma_0, \label{EQ1}\\
&e^{\Theta'}e^{\tilde{\Theta'}} = 1, \label{EQ2}
\end{align}
where we restrict our discussion to gauge transformations continuously connected to identity. 
General solution of these equations includes $\Theta \sim so(4,4)$, which is a linear combination of 28 gauge transformation generators
\begin{equation}
\{\gamma_a, \gamma_a\gamma_b, \Gamma_a\Gamma_b, i\Gamma_j, \Gamma_0\gamma_j\Gamma_k ; j, k = 1,2,3, a, b = 0,1,2,3, a > b \} \in \Theta, \label{LROT}
\end{equation}
and $\Theta' \sim sp(8)$, which is a linear combination of 36 gauge transformation generators of pseudoscalar, all bivectors, and all trivectors
\begin{equation}
\{i, \gamma_j\Gamma_k, \gamma_k\gamma_l, \Gamma_k\Gamma_l, \gamma_0, \Gamma_0, \gamma_0\gamma_j\Gamma_k, \Gamma_0\gamma_j\Gamma_k; j, k, l= 1,2,3, k > l \} \in \Theta'.
\end{equation}
The de Sitter algebra $\Theta_{DS} \sim so(1,4)$
\begin{equation}
\{\gamma_a, \gamma_a\gamma_b\} \in \Theta_{DS}
\end{equation}
is a subalgebra of $\Theta$. 

The Clifford odd parts of $\Theta$ and $\Theta'$ mix odd (left-handed $\psi_L$) and even (right-handed $\psi_R$) spinors. Since we know that left- and right-handed spinors transform differentially, only Clifford even subalgebras of $\Theta$ and $\Theta'$ are permitted, namely
\begin{align}
&\{\gamma_a\gamma_b, \Gamma_a\Gamma_b\} \in \Theta_{Even} \sim so(1,3) \oplus so(1,3), \label{LROT2} \\
&\{i, \gamma_j\Gamma_k, \gamma_k\gamma_l, \Gamma_k\Gamma_l \} \in \Theta'_{Even} \sim u(1) \oplus  so(6) \sim u(1) \oplus su(4).
\end{align}

The gauge transformations $\{\Gamma_a\Gamma_b\}$ can be further decomposed into weak transformations $\{\Gamma_k\Gamma_l\}$ and weak boost transformations $\{\Gamma_0\Gamma_j\}$, which are counterparts of spacial rotation $\{\gamma_k\gamma_l\}$  and Lorentz boost transformations $\{\gamma_0\gamma_j\}$.

Unitary algebra $u(3)$ is embedded in $\{\gamma_j\Gamma_k, \gamma_k\gamma_l, \Gamma_k\Gamma_l \} \sim su(4)$.  Removing $u(1)$ $\{J\}$ from $u(3)$ defines the color algebra $su(3)$. 

Since there are left-handed weak $su(2)_L$ and right-handed weak $u(1)_R$, one might expect left-right symmetric $su(2)_R$ as well. We can even go further and entertain the possibility of two exact copies of left-handed $\Theta_{EvenL}$ and right-handed $\Theta_{EvenR}$. It means that left- and right-handed fermions have separate local Lorentz gauge symmetries $\{\gamma_a\gamma_b\}_L$ and $\{\gamma_a\gamma_b\}_R$. Thus there are left- and right-handed gravities.

Of course, the symmetries studied in this section are speculative in nature. 
If there is indeed grand unification scale physics involving $\Theta_{EvenL}$, $\Theta_{EvenR}$ and $\Theta_{Even}'$, either symmetry breaking or other mechanism is needed to prevent detection of gauge interactions related to pseudoscalar $\{i\}$, quark/lepton mixing part of $su(4)$, weak boost $\{\Gamma_0\Gamma_j\}$, $W_R^{\pm}$ part of $su(2)_R$, and differences between left-handed $\{\gamma_a\gamma_b\}_L$ and right-handed $\{\gamma_a\gamma_b\}_R$ Lorentz transformations. It's an interesting topic. Nevertheless, we leave it to future research.

\section{Conclusion}
We propose a model which is based on Clifford algebra $C\!\ell_{0,6} \oplus C\!\ell_{T1}$. With the purpose of studying 3 generations of standard model fermions, a ternary Clifford vector is introduced alongside 6 binary Clifford vectors. The model includes local gauge symmetries $SO(1,3)_{LOR} \otimes SU(2)_{WL} \otimes U(1)_{WR} \otimes U(1)_{B-L} \otimes SU(3)_C$.

There are two sectors of bosonic fields as electroweak and Majorana bosons. Electroweak boson field interacts with both left-handed and right-handed fermions. Majorana boson field interacts with right-handed fermions only. 

The electroweak boson sector is composed of scalar, pseudoscalar, and antisymmetric tensor components. Scalar and/or pseudoscalar Higgs fields break the electroweak symmetry, contributing masses to fermions. 

The Majorana boson sector is comprised of neutrino Majorana Higgs and pseudo-Nambu-Goldstone bosons. It is responsible for flavor mixing between generations. The neutrino Higgs field part of Majorana boson sector acquires a nonzero VEV via spontaneous symmetry breaking, inducing Majorana masses of right-handed neutrinos.  

The LHC 750 GeV diphoton resonance might possibly be identified as a Majorana sector pseudo-Nambu-Goldstone boson (pNGB), which results from spontaneous symmetry breaking of a flavor-related global $U(1)$ symmetry. The symmetry involves right-handed leptons and quarks. The pNGB is not exactly massless, since the electroweak sector explicitly breaks the global symmetry and can generate mass for it. The diphoton decay of the pseudo-Nambu-Goldstone boson is loop induced, since tree-level decay is suppressed by large Majorana mass of the right-handed neutrino. 

The pNGB might be a composite boson representing right-handed-four-fermion condensation $\bar{\nu}_{e R}b_R\bar{t}_{R}e_R$ (and/or $\bar{u}_R\tau_R\bar{\nu}_{\tau R}d_R$, $\bar{c}_R\mu_R\bar{\nu}_{\mu R}s_R$) via dynamical symmetry breaking. Together with two other configurations $\bar{\nu}_{\mu R}\tau_R\bar{\nu}_{\tau R}\mu_R$ and $\bar{u}_Rs_R\bar{c}_Rd_R$, four-fermion condensations are also potential dark matter candidates, since they are free from direct gauge interactions. The prime dark matter candidate might be the four-lepton condensation $\bar{\nu}_{\mu R}\tau_R\bar{\nu}_{\tau R}\mu_R$.

\section*{Acknowledgments}

I am grateful to Norma Susana Mankoc Borstnik, Salvatore Capozziello, Diego Julio Cirilo-Lombardo, Matej Pavsic, and Hidezumi Terazawa for helpful correspondences.

\end{document}